\documentclass[prl,aps,twocolumn,showpacs,superscriptaddress,amsmath,amssymb,floatfix]{revtex4}
\usepackage{graphicx}

\newcommand{\be}{\begin{equation}}
\newcommand{\ee}{\end{equation}}

\newcommand{\rr}{{\mathbf r}}

\newcommand{\eq}[1]{(\ref{#1})}

\begin{document}
\title{Synchronized and desynchronized phases of coupled non-equilibrium exciton-polariton condensates}
\author{Michiel Wouters}
\affiliation{Institute of Theoretical Physics, Ecole Polytechnique
F\'ed\'erale de Lausanne EPFL, CH-1015 Lausanne, Switzerland}
\affiliation{TFVS, Universiteit Antwerpen, Groenenborgerlaan 171,
2020 Antwerpen, Belgium}
\begin{abstract}
We theoretically analyze the synchronized and desynchronized phases of coupled non-equilibrium polariton condensates within mean field theory. An analytical condition for the existence of a synchronized phase is derived for two coupled wells. The case of many wells in a 2D disordered geometry is studied numerically. The relation to recent experiments on polariton condensation in CdTe microcavities is discussed.
\end{abstract}
\maketitle


Polariton condensates in semiconductor microcavities provide us with novel macroscopic quantum objects, whose precise experimental \cite{expt_refs} and theoretical \cite{nonresonant,marzena,keeling,boseglass,davide} characterization is presently an active field of research. A realistic description of the actually realized polariton condensates requires to take into account the interactions between polaritons, pumping, losses and the disordered potential landscape. This potential is due to fluctuations in the growth process of the semiconductor heterostructure and contains in CdTe microcavities typically a few deep minima within the condensate area. In the presence of such a potential, the question naturally arises whether a single condensate is formed that is modulated by the disorder potential yet phase-coherent over its size, or rather several independent condensates with different frequencies are formed. In recent experiments~\cite{augustin_coh}, it was found that depending on the configuration of the external potential (position on the sample), both the case of a fully coherent condensate with a single frequency (synchronized) and the case of incoherent condensates with different frequencies (desynchronized) can be realized. The experimental data suggest that the underlying physics is analogous to the technologically important phenomenon of mode locking in laser gyroscopes \cite{meystre}. 

In the present paper, we will present a theoretical picture of the mode synchronization in polariton condensates. Because the average frequency of the polariton condensate is related to the macroscopically occupied state, a mean field description can be used in a first approximation. A non-equilibrium mean field model for polariton condensates was proposed in Ref. \cite{nonresonant} and used to describe their peculiar spatial and spectral shape in Ref. \cite{shape}. A similar model was introduced in Ref. \cite{keeling}. Within the latter framework, mode locking effects between discrete energy levels polariton condensates are investigated by P. Eastham \cite{paul_sync}.

The simplest system that allows to understand the synchronization physics of a polariton condensate in a disordered microcavity consists of two coupled wells with a potential difference. Our analysis will show that a Josephson flow is responsible for the phase locking between the condensates. If the potential difference between the wells exceeds a certain critical value, the Josephson flow cannot reach a steady state and the condensates no longer share the same frequency. Two condensates with different frequencies are formed in each well and density oscillations reminiscent of the AC Josephson effect appear.

The dynamics of the condensate macroscopic wavefunction $\psi(\rr)$ will be described by a generalized Gross-Pitaevskii equation~\cite{nonresonant}. Following the work on Josephson physics with spatially separated ultracold atomic Bose-Einstein condensates~\cite{oberthaler,bec-book} , the condensate wave function is projected on the wave functions  $\phi_{1,2}$ of each well (normalized as usual as $\int\!d\mathbf{r}\,|\phi_j|^2=1$). In terms of the amplitudes $\psi_{1,2}$ in the two wells, the total polariton wavefunction reads
$
\psi(\mathbf{r})=\psi_{1}\,\phi_1(\mathbf{r})+\psi_2\,\phi_2(\mathbf{r})
$
and the dynamics is given by~\cite{nonresonant}
\begin{eqnarray}
i \frac{d\, \psi_{j}}{d t}\!\!\!&=&\!\!\! \epsilon_j \psi_j -J\,\psi_{3-j}+
(U_j\,|\psi_{j}|^2 + U^R_j n_j)\,\psi_{j} \nonumber \\
&&+\frac{i}{2}\big[R(n_j)-\gamma \big]\psi_j  \label{JosephsonPsi},
\end{eqnarray}
where $\gamma$ is the polariton line width (for simplicity taken to be equal in both wells), $J$ is the hopping energy and $\epsilon_j$ is the ground state energy of each well in the absence of coupling. The term $R(n_j)$ describes the gain of the condensate due to the stimulated scattering from the excitonic reservoir into the lower polariton states.
Polariton-polariton interactions are described by the charging energy $U_j |\psi_j|^2$ and the interactions between the reservoir and condensate polaritons are taken into account by the term $U^R_j n_j$.

The reservoir occupation $n_{1,2}$  results from the equilibrium between the pumping at a rate $P_{1,2}$ and its decay:
\begin{equation}
	\frac{d\, n_{j}}{dt}=P_j-\gamma_R\,n_j-R(n_j)|\psi_j|^2,
	\label{JosephsonN}
\end{equation}
where the term $\gamma_R n_j$ represents losses through channels different from the condensate. The stationary state and excitation spectrum of the symmetric system ($\epsilon_1=\epsilon_2$, $U_1=U_2$ and $U_1^R=U_2^R$) are discussed in Ref.~\cite{nonresonant}, where it was found that the Josephson plasma oscillations are damped due to the non-equilibrium nature of the polariton condensate.

\begin{figure*}[htb]
\begin{center}
$\begin{array}{c@{\hspace{0.2in}}c}
\includegraphics[width=0.8 \columnwidth,angle=0,clip]{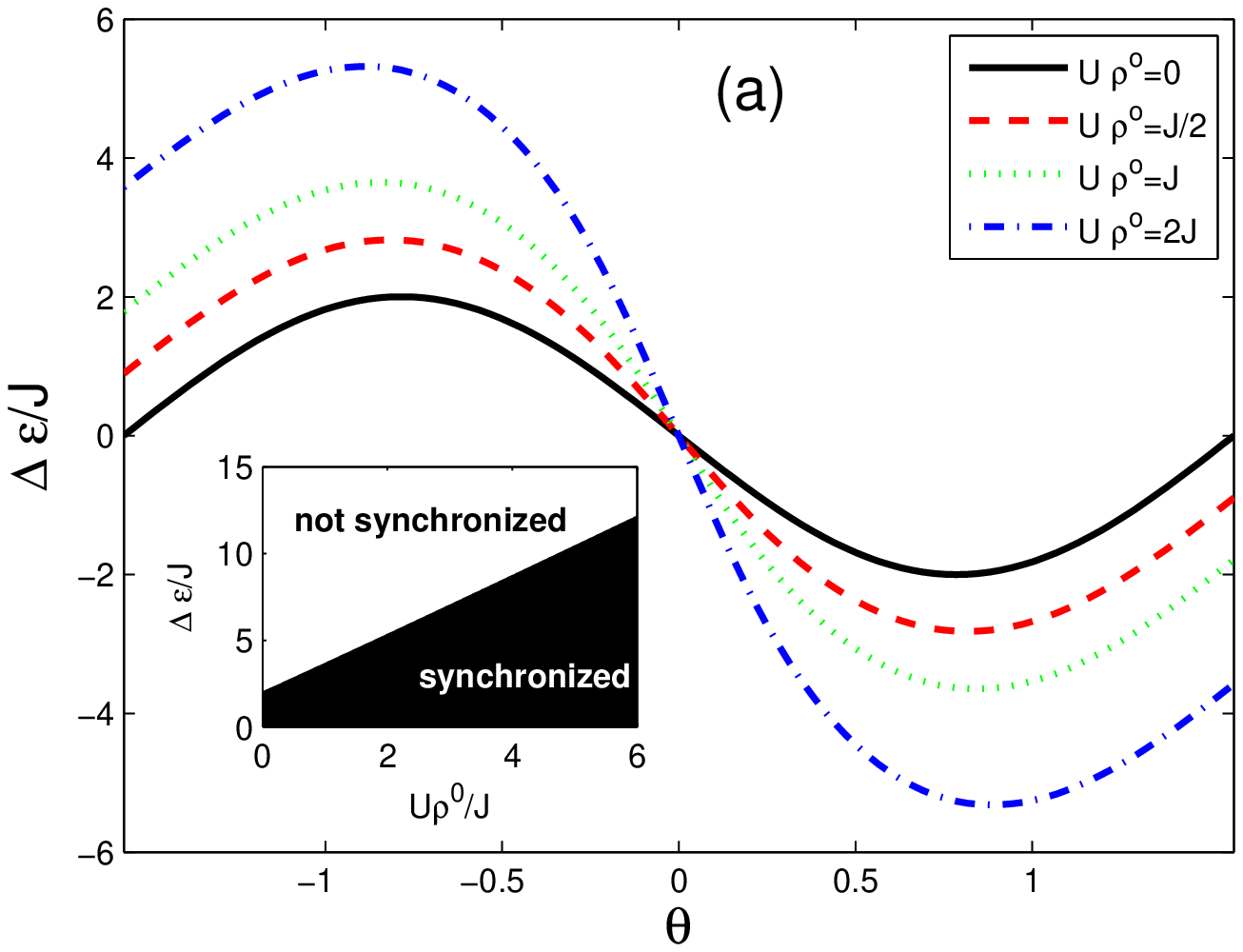} &
\includegraphics[width=0.8 \columnwidth,angle=0,clip]{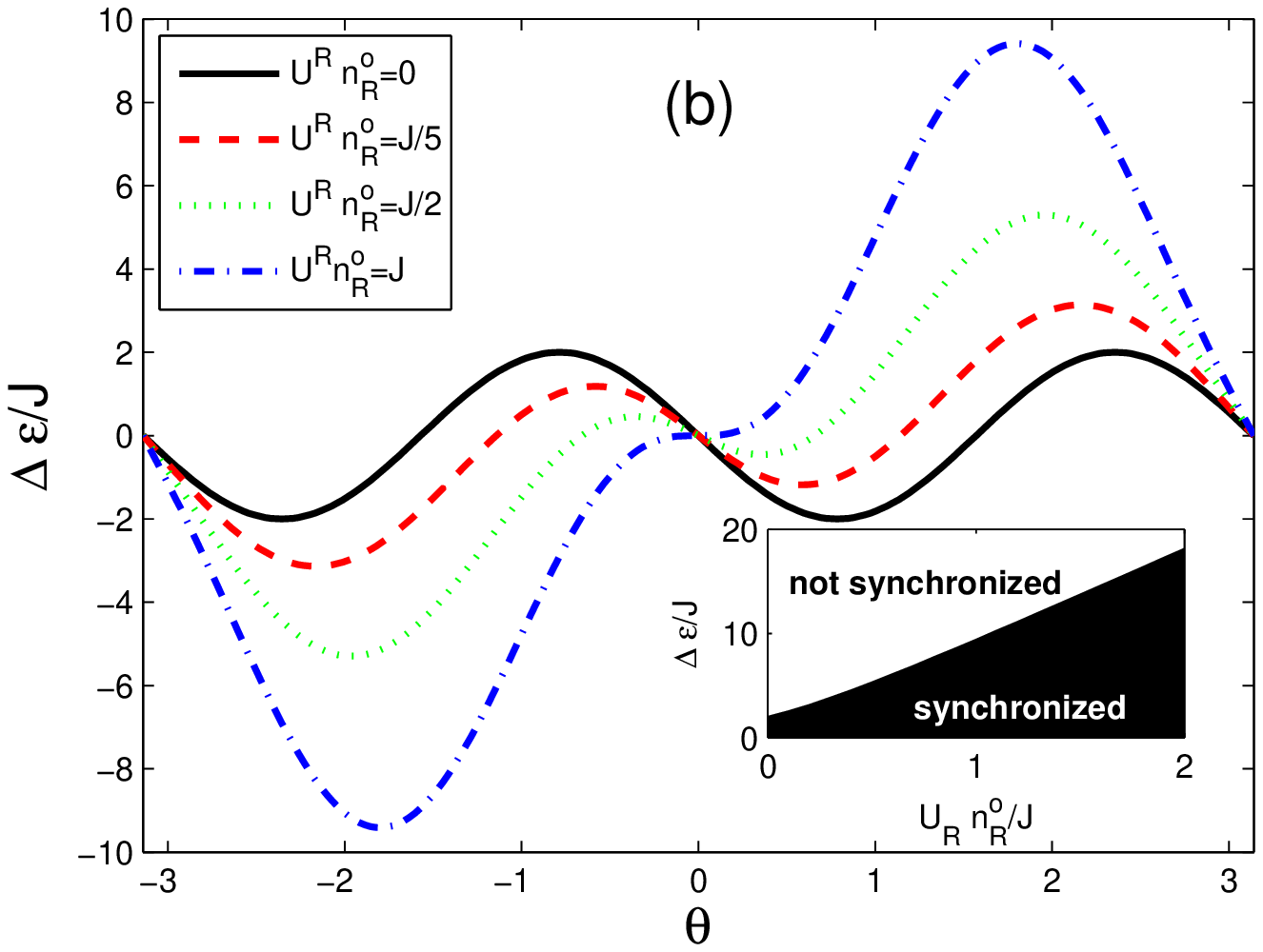}
\end{array}$
\end{center}
\caption{(Color Online) (a) The effective detuning $\Delta \epsilon$ (in units of the tunneling rate $J$) as a function of the phase difference $\theta$. The damping rates are taken $\gamma/J=\gamma_R/J=1$  and equal pump rates $P_1=P_2$. The blue shifts in the absence of the coupling $U\rho^o$ are shown in the legend and $U^R_{1,2}=0$. The inset shows the maximum detuning as a function of blue shift in the absence of coupling.
(b) The same for zero condensate-condensate interaction $U_{1,2}=0$ and non-vanishing reservoir-condensate interactions $U_R^1=U_R^2=U_R$ (see legend). The inset shows the maximum detuning as a function of the blue shift in the absence of coupling $U_R n_R^o$.}
\label{fig:two_an}
\end{figure*}

Let us start to analyze under which conditions a synchronized steady state exists if the two potential wells are detuned. The wave functions in the two wells then take the form $\psi_{1,2}(t)=\sqrt{\rho_{1,2}}e^{ i(-\omega t\pm\frac\theta 2)}$ and $n_j(t)=n_j$. Substituting this state in the equations \eq{JosephsonPsi} and \eq{JosephsonN} gives
\begin{eqnarray}
	&&\omega \sqrt{\rho_{j}} = (\epsilon_{j}+ U_{j} \rho_{j} + U^R_{j} n_{j} )\sqrt{\rho_{j}} \nonumber \\
	&& -J \sqrt{\rho_{3-j}}\exp[-i (-1)^j\theta]+\frac{i}{2} [R(n_{j})-\gamma] \sqrt{\rho_{j}}    \label{eq:statpsi}  \\
	&& P_{j}=\gamma_R n_{j}+R(n_{j}) \rho_{j}.
	\label{eq:statN}
\end{eqnarray}
The imaginary part of Eq. \eq{eq:statpsi}, expressing the conservation of polariton density
\begin{equation}
	[\gamma-R(n_{1,2})] \rho_{1,2} = \pm 2 J \sqrt{\rho_1 \rho_2} \sin(\theta),
	\label{eq:statim}
\end{equation}
shows that a Josephson current $I_{J}=2 J \sqrt{\rho_1 \rho_2} \sin(\theta)$ flows between the two condensates. With the help of Eqns. \eq{eq:statpsi} and \eq{eq:statim}, the recombination rate $R(n_{1,2})$ can be written as
\begin{equation}
	R(n_{1,2})=\gamma-\big[ \Delta \omega \tan \theta
	\mp \sqrt{(\Delta \omega \tan \theta)^2 + 4 J^2 \sin^2 \theta} \; \big],
	\label{eq:statR}
\end{equation}
where the effective detuning $\Delta \omega$ is shifted from the bare detuning $\Delta \epsilon=\epsilon_1-\epsilon_2$ by the mean field shifts:
\begin{equation}
	\Delta \omega = \Delta \epsilon +U_1 \rho_1 +U^R_1 n_1 - U_2 \rho_2 - U^R_2 n_2.
	\label{eq:dom_deps}
\end{equation}
Eq. \eq{eq:statR} together with Eq. \eq{eq:statN}
finally allows to find the phase difference $\theta$ as a function of the effective condensate detuning $\Delta \omega$.

A simple analytical expression for $\Delta \omega$ as a function of the phase difference $\theta$ exists for $P_1=P_2=P$ and when the decay of reservoir polaritons is most efficient through the condensate: $R(n_j) \rho_j \gg \gamma_R n_R$, a condition that is expected to hold well above the threshold. The solution to Eqns. (\ref{eq:statN}) and (\ref{eq:statR})  then reads
\begin{equation}
	\Delta \omega = - \frac{2J^2}{\gamma} \sin(2\theta),\;\;\; {\rm if}\; R(n_j) \rho_j \gg \gamma_R n_R.
	\label{eq:delta_omega}
\end{equation}
As expected, the maximal value of the detuning $\Delta \omega_c = 2J^2/\gamma$ increases as a function of the coupling between the wells $J$ and is inversely proportional to the polariton line width $\gamma$.

Under the condition that the nonlinearity $U_j$ in both wells is equal, also a simple analytical expression for the bare detuning $\Delta \epsilon$ [see Eq.\eq{eq:dom_deps}] as a function of the phase difference can be derived.
Some algebra with Eqns. \eq{eq:statpsi} to \eq{eq:delta_omega} yields
\begin{multline}
\Delta \epsilon = -\frac{2J^2}{\gamma} \sin(2\theta)
-\frac{4J U_1 \rho^o}{\gamma}\frac{\sin \theta}{\sqrt{1+(2J/\gamma)^2 \sin^2 \theta}}\\
-U^R_1 n_1(\theta) + U^R_2 n_2(\theta),
\label{eq:delta_eps}
\end{multline}
where $\rho^o$ is the condensate density in both wells in the absence of coupling ($J=0$) and $n_{1,2}(\theta)$ is defined by Eqs. \eq{eq:statR} and \eq{eq:delta_omega}. Note that for vanishing tunneling rate $J$, a synchronized solution only exists for $\Delta \epsilon = 0$: local interactions alone cannot lock the two spatially separated condensates to a single frequency.

A plot of the detuning as a function of the phase difference between the two wells according to Eq. \eq{eq:delta_eps} is shown in Fig. \ref{fig:two_an} (a) for several values of $U_1\rho^o$ and zero interaction strength between the reservoir and condensate $U^R_{1,2}=0$. Up to a critical value of the detuning, a stationary state with a single frequency for the two condensates exists. The detuning reaches a maximum value $\Delta \epsilon_{\rm max}$ for a phase difference $\theta \approx \pi/4$, at which the flow $I_{J}=2 J \sqrt{\rho_1 \rho_2} \sin(\theta)$is maximal. For a detuning larger than $\Delta \epsilon_{\rm max}$ no stationary synchronized solution exists. It is important to note that the stationary synchronized state does not coincide with a linear eigenstate of the two-well system, for which $\theta\in \{0,\pi\}$, but that the condensation occurs in a new state that is formed above the threshold for condensation.

The existence of a synchronized state up to a critical detuning is very similar to the phenomenon of mode locking in lasers that is most simply described by the Adler equation for the phase difference $\theta$ between two modes \cite{meystre}. The most important difference between polariton condensates and ordinary lasers is the large value of the polariton nonlinearity as compared to the very small photon nonlinearity in ordinary lasers. Fig. \ref{fig:two_an} (a) shows that the polariton interactions help the synchronization of the polariton condensate. The physical reason is that the flow depletes the high energy well. As a consequence,  the blue shift of the higher well decreases and the energy levels are pulled toward each other. The phase diagram as a function of the detuning $\Delta \epsilon$ and pump power (expressed in terms of $\rho^o$) is shown in the inset of Fig. \ref{fig:two_an} (a). 

The interactions between the reservoir and condensate polaritons can enhance the synchronization as well [see Fig. \ref{fig:two_an} (b)].  Where the condensate-reservoir interactions counteract synchronization at small phase difference, a strong enhancement is obtained for $\theta$ close to $\pi/2$. Within the approximation that $\gamma_R=0$, the contribution to the detuning from condensate-reservoir interactions does not depend on the pump intensity. The experimental study of the synchronization transition as a function of the pump power could therefore give an indication whether rather the condensate-condensate or condensate-reservoir interactions are the dominant mechanism for the synchronization. Unfortunately, the independence of the pump intensity is spoiled when the finite line width of the reservoir excitons $\gamma_R$ is taken into account. An alternative physical quantity that could give an indication about the dominant mechanism for the frequency synchronization of the condensates is the relative phase between the two condensates. A numerical stability analysis of the synchronized state and a full integration of the motion equations \eq{JosephsonPsi} and \eq{JosephsonN} have shown that the expected window for the relative phase when condensate-condensate interactions prevail is $0<|\theta|<\pi/4$, where $\pi/2<|\theta|<\pi$ when condensate-reservoir interactions dominate.


\begin{figure}[htbp]
\begin{center}
\includegraphics[width=\columnwidth,angle=0,clip]{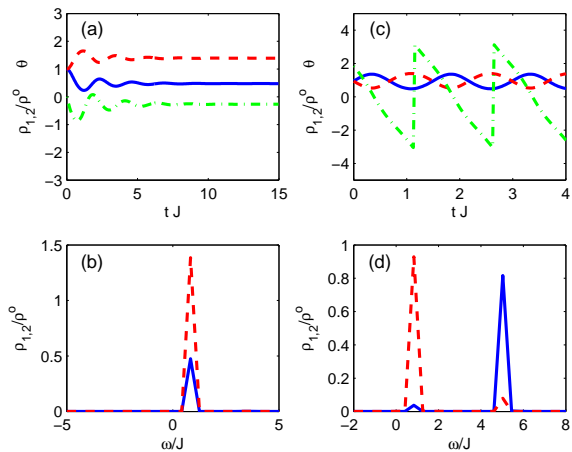}
\end{center}
\caption{(Color Online) Time evolution of the density (full ($\rho_1$) and dashed ($\rho_2$) lines) and phase difference (dash-dotted line) in a two-well geometry with frequency detuning $\Delta \epsilon / J =2$ (a) and $\Delta \epsilon / J =3.8$ (c). (b) and (d) show the temporal Fourier transforms. Other parameters: $\gamma/J=1,\; \gamma_R/J=5$ and $UP/J^2=1$. }
\label{fig:twowell}
\end{figure}


In agreement with the analytical prediction, the numerical integration of Eqns. \eq{JosephsonPsi} and \eq{JosephsonN} with the parameters for Fig. \ref{fig:twowell}a,b shows a synchronized solution. The densities evolve to a constant value that is symmetric around the density in the absence of coupling $\rho=\rho^o$. The condensate at higher energy (full line) is less occupied due to the flow into the lower energy one (dashed line). The temporal Fourier transform (Fig. \ref{fig:twowell}b) is peaked at a single frequency for both condensates.

In Fig. \ref{fig:twowell}c,d, the ratio of the frequency detuning with respect to the coupling $\Delta \epsilon/J = 3.8$ is too large to allow for the locking of the phases in the two wells. The time dependent phase difference induces an oscillating Josephson current that bears a striking analogy with the AC Josephson effect, where the application of a constant voltage (chemical potential difference) also leads to an alternating current. A crucial difference with the AC Josephson effect in atomic Bose-Einstein condensates concerns the relaxation: where the AC Josephson oscillations of an atomic condensate are damped at any finite temperature and its steady state is at thermodynamical equilibrium with a single chemical potential, no relaxation to a single frequency state is possible for a desynchronized non-equilibrium polariton condensate.

In the frequency domain, the alternating Josephson currents cause each condensate to have a small contribution at the frequency of the other condensate (see Fig. \ref{fig:twowell}d). These overlapping frequency components give rise to residual coherence and thus interference fringes. So far, this residual coherence was not observed experimentally~\cite{augustin_coh} in the desynchronized regime. To verify our prediction of residual coherence and density oscillations in the desynchronized regime, the experimental investigations could be performed in a more controlled way by using polariton condensates in mesa microcavities~\cite{plots}.

It should also be mentioned that for values of the detuning, close to the maximal value $\Delta \epsilon \lesssim \Delta \epsilon_c$, numerical integration of the Josephson equations shows that both the solutions with stationary and oscillating densities can be reached, depending on the initial conditions.


\begin{figure}[htbp]
\begin{center}
\includegraphics[width=1\columnwidth,angle=0,clip]{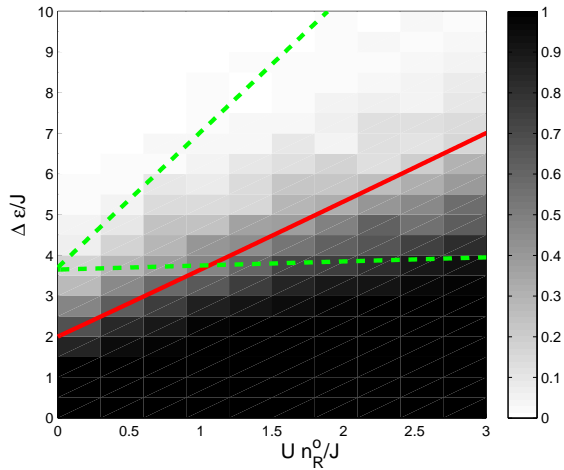}
\end{center}
\caption{(Color Online) The probability to have a synchronized condensate in a 2D disorder potential with uniformly distributed energy levels in the interval $[0,\; \Delta \epsilon]$ is shown in gray scale. Only interactions within the condensate are considered ($U^R_{1,2}=0$). The full line shows the boundary of the locked solution for two wells that are detuned by an energy $\Delta \epsilon$. The dashed lines show the border of synchronization when only the central well in the 2D geometry is detuned by an energy $+\Delta \epsilon$ (lower dashed line) or $-\Delta \epsilon$ (upper dashed line). From a a 5x5 square of coupled wells, the central 3x3 part is uniformly excited by the pump laser.}
\label{fig:disorder_pd}
\end{figure}

The generalization of the model equations \eq{JosephsonPsi} and \eq{JosephsonN} to the case of multiple wells with randomly distributed energy levels, closer to the disordered reality of CdTe microcavities, is straightforward. 
Fig. \ref{fig:disorder_pd} shows the probability to reach a synchronized state as a function of the disorder strength and interactions. The averaging of the synchronization over many realizations of the disorder potential gives rise to a transition region in the phase diagram where the synchronization depends on the actual realization of the disordered potential. As a general trend, the interactions increase the probability to reach a synchronized state as well as the width of the transition region. For comparison, the analytically determined phase boundary of the two-level system \eq{eq:delta_omega} is shown by the full line. 
The dashed lines show the phase boundary for a regular 2D array of energy levels where the central level is detuned by an energy $+\Delta \epsilon$ (lower dashed line) and by an energy $-\Delta \epsilon$ (upper dashed line). 
These lines show that a higher dimensionality favors the synchronization for the non-interacting polariton condensate. A physical explanation for this fact is that the total Josephson flow out of (or into) the central well can reach a higher value because it is distributed over more links. Another clear feature is that the interactions help the synchronization much more if the central well is negatively detuned. 
Note finally that the presently considered inherently non-equilibrium desynchronized phase is very different from the thermodynamic equilibrium insulating phase studied for microcavity polaritons in Ref. \cite{boseglass}.


In summary, we have analyzed the recently observed synchronization-desynchronization transition of polariton condensates within mean field theory. An analytical condition for synchronization is derived for the case of a two-well system. The many well configuration was analyzed numerically. The same phenomenology was found in both cases: for small detuning between the different wells, the Josephson currents and densities reach a steady state. For too large detuning on the other hand, a steady state no longer exists and the Josephson currents cause density oscillations in the different condensates. Both interactions of the condensate with itself and with the reservoir are shown to enhance the synchronization. The similarities and differences with the Josephson effect of superfluids and the locking-delocking transition in ordinary lasers was clarified. 

I am grateful to D. Sarchi, V. Savona and I. Carusotto for continuous stimulating discussions and to I. Carusotto also for his comments on the manuscript. I wish to thank M. Richard, A. Baas, K. Lagoudakis for many fruitful discussions on their experiments and P. Eastham for sharing his insights in the synchronization of polariton condensates.

\end{document}